\documentclass[tradiabstract]{aa}  

\usepackage{graphicx}
\usepackage[dvipsnames]{xcolor}
\usepackage{txfonts}
\usepackage{appendix}
\usepackage{hyperref}
\usepackage{amssymb}

\begin{document} 

\newcommand{\Ho}{\mbox{$H_0$}}
\newcommand{\ang}{\mbox{{\rm \AA}}}
\newcommand{\abs}[1]{\left| #1 \right|} 
\newcommand{\avg}[1]{\left\langle #1 \right\rangle} 
\newcommand{\kms}{\ensuremath{{\rm km\,s^{-1}}}}
\newcommand{\zabs}{\ensuremath{z_{\rm abs}}}
\newcommand{\zem}{\ensuremath{z_{\rm quasar}}}
\newcommand{\cmsq}{\ensuremath{{\rm cm}^{-2}}}
\newcommand{\ergs}{\ensuremath{{\rm erg\,s^{-1}}}}
\newcommand{\ergsa}{\ensuremath{{\rm erg\,s^{-1}\,{\AA}^{-1}}}}
\newcommand{\ergscm}{\ensuremath{{\rm erg\,s^{-1}\,cm^{-2}}}}
\newcommand{\ergscma}{\ensuremath{{\rm erg\,s^{-1}\,cm^{-2}\,{\AA}^{-1}}}}
\newcommand{\msyr}{\ensuremath{{\rm M_{\rm \odot}\,yr^{-1}}}}
\newcommand{\nhi}{n_{\rm HI}}
\newcommand{\fhi}{\ensuremath{f_{\rm HI}(N,\chi)}}
\newcommand{\Av}{\ensuremath{A_V}}
\newcommand{\lya}{Ly-$\alpha$}
\newcommand{\Hb}{H-$\beta$}
\newcommand{\OVI}{\ion{O}{vi}}
\newcommand{\OIII}{\ion{O}{iii}}
\newcommand{\OII}{\ion{O}{ii}}
\newcommand{\OI}{\ion{O}{i}}
\newcommand{\HI}{\ion{H}{i}}
\newcommand{\HeII}{\ion{He}{ii}}
\newcommand{\HH}{\ensuremath{{\rm H}_2}}
\newcommand{\SII}{\ion{S}{ii}}
\newcommand{\SiIII}{\ion{Si}{iii}}
\newcommand{\SiIV}{\ion{Si}{iv}}
\newcommand{\SiII}{\ion{Si}{ii}}
\newcommand{\AlIII}{\ion{Al}{iii}}
\newcommand{\AlII}{\ion{Al}{ii}}
\newcommand{\ArI}{\ion{Ar}{i}}
\newcommand{\FeII}{\ion{Fe}{ii}}
\newcommand{\ZnII}{\ion{Zn}{ii}}
\newcommand{\CrII}{\ion{Cr}{ii}}
\newcommand{\MnII}{\ion{Mn}{ii}}
\newcommand{\MgII}{\ion{Mg}{ii}}
\newcommand{\MgI}{\ion{Mg}{i}}
\newcommand{\NiII}{\ion{Ni}{ii}}
\newcommand{\NV}{\ion{N}{v}}
\newcommand{\CIV}{\ion{C}{iv}}
\newcommand{\CIII}{\ion{C}{iii}}
\newcommand{\CII}{\ion{C}{ii}}
\newcommand{\CI}{\ion{C}{i}}
\newcommand{\CaII}{\ion{Ca}{ii}}
\newcommand{\TiII}{\ion{Ti}{ii}}
\newcommand{\JB}{J\,0007-5705}
\newcommand{\JBlong}{J\,000736.56-570151.8}

\newcommand{\iap}{Institut d'Astrophysique de Paris, CNRS-SU, UMR\,7095, 98bis bd Arago, 75014 Paris, France -- \email{noterdaeme@iap.fr}\label{iap}}
\newcommand{\ioffe}{Ioffe Institute, {Polyteknicheskaya 26}, 194021 Saint-Petersburg, Russia -- \email{s.balashev@gmail.com}\label{ioffe}}
\newcommand{\oat}{INAF - Osservatorio Astronomico di Trieste, Via G.B. Tiepolo, 11, 34143 Trieste, Italy \label{oat}}
\newcommand{\ifpu}{IFPU - Institute for Fundamental Physics of the Universe, via Beirut 2, 34151 Trieste, Italy \label{ifpu}
}
\newcommand{\infn}{National Institute for Nuclear Physics, via Valerio 2, 34127 Trieste, Italy \label{infn}}
\newcommand{\op}{LERMA, Observatoire de Paris, Université PSL, Sorbonne Université, 75014 Paris, France \label{op}}
\newcommand{\ens}{Laboratoire de Physique de l’Ecole Normale Supérieure, PSL, CNRS, SU, Université de Paris, 75005 Paris, France \label{ens}}
\newcommand{\victoria}{NRC Herzberg Astronomy and Astrophysics Research Centre, 5071 West Saanich Road, Victoria, BC V9E 2E7, Canada \label{victoria}}
\newcommand{\uchile}{Departamento de Astronomía, Universidad de Chile, Casilla 36-D, Santiago 7550000, Chile \label{uchile}}
\newcommand{\durham}{Centre for Extragalactic Astronomy, Durham University, South Road, Durham DH1 3LE, UK \label{durham}}

        \title{One H$_2$ molecule per ten million H-atoms reveals  
        sub-pc scale cold overdensities at $z\sim4$\thanks{Based on observations collected at the European Southern Observatory under programme 112.25NR.}
        }
                 
 \author{
   P.~Noterdaeme\inst{\ref{iap}}
  \and
  S.~Balashev\inst{\ref{ioffe}} 
  \and 
  T.~Berg\inst{\ref{victoria}} \and
  S.~Cristiani\inst{\ref{oat},\ref{ifpu},\ref{infn}} \and
  R.~Cuellar\inst{\ref{uchile}} \and
  G. Cupani\inst{\ref{oat}} \and
  S.~Di~Stefano\inst{\ref{oat}} \and
  V.~D'Odorico\inst{\ref{oat}} \and
  C.~Fian\inst{\ref{oat}} \and
  B.~Godard\inst{\ref{op},\ref{ens}} \and
  S. L{\'o}pez\inst{\ref{uchile}} \and
  D.~Milakovi\'c\inst{\ref{oat},\ref{ifpu}} \and
  A.~Trost\inst{\ref{oat}} \and
  L. Welsh\inst{\ref{oat},\ref{durham}} 
   }
   \institute{\iap \and \ioffe \and \victoria \and \oat \and \ifpu \and 
   \infn \and 
   \uchile \and \op \and \ens \and \durham}
     \authorrunning{P.~Noterdaeme et al.}        
   \date{\today.}

    \abstract{
We present the detection and analysis of H$_2$ absorption at $z = 4.24$ towards the bright quasar \JB, observed with the Very Large Telescope as part of the ESPRESSO QUasar Absorption Line Survey (EQUALS). The high resolving power, $R \approx 120\,000$, enables the identification of extremely weak H$_2$ lines in several rotational levels at a total column density of $N({\rm H_2}) \approx 2\times10^{14}\,\mathrm{cm^{-2}}$, among the lowest ever measured in quasar absorption systems. Remarkably, this constitutes the highest-redshift H$_2$ detection to date.

Two velocity components are resolved, separated by only 3\,\kms: a narrow ($b \sim 1.7$\,\kms) and a broader ($b \simeq 6.2$\,\kms) component. Modelling the rotational population of H$_2$ yields density of $\log n_{\rm H}/$cm$^{-3} \sim 2.8$ with temperature of $\sim$40\,K (typical of the cold neutral medium) for the narrow component and $\log n_{\rm H}/$cm$^{-3} \sim 1.4$ , $T\sim600$\,K for the warmer, more turbulent component under a moderate ultraviolet (UV) field, suggesting at least several Mpc distance from the quasar. 

This system reveals the existence of tiny (down to $\sim 0.01$ pc), cold {overdensities} in the neutral medium. 
Their detection among only 7 damped Lyman-$\alpha$ systems in EQUALS suggests that they may be widespread yet usually remain undetected. H$_2$ provides an exceptionally sensitive probe of these structures: even a minute molecular fraction produces measurable Lyman-Werner absorption lines along the extremely narrow optical beam --the size of the quasar's accretion disc-- when observed at sufficiently high spectral resolution. High-resolution spectroscopy on extremely large telescopes may routinely detect and resolve such structures in the distant Universe, when 21-cm absorption will trace the collective contribution of many {cold} cloudlets toward larger radio background sources.
    }
  

\maketitle

\section{Introduction}
{Molecular hydrogen (H$_2$), the simplest and by far the most abundant molecule in the Universe, plays a central role in tracing the physical conditions of the interstellar and circumgalactic medium. }
Owing to its specific formation, destruction, and excitation processes, 
it provides sensitive diagnostics of gas density, temperature, and radiation field 
{via its rotational and vibrational populations} \citep[e.g.,][]{Jura1975,Bialy2017}.
 In addition, 
 H$_2$ can reveal dynamical effects such as 
dissipation of turbulent energy in vortices and shocks \citep[e.g.,][]{Godard2009,Lesaffre2020} and phase transition processes 
such as condensation and evaporation of cold neutral medium (CNM) clouds \citep{Valdivia2016,Bellomi2020}.

{Infrared H$_2$ emission lines are typically detected in dense warm gas illuminated by massive stars \citep{Habart2003} but can also arise from diffuse and translucent gas, likely due to reprocessing of mechanical or ultraviolet (UV) energy \citep{Villa-Velez2024}.
Nevertheless,} 
the most sensitive way to detect H$_2$ remains 
through electronic (Lyman and Werner) absorption lines against background sources. 
Since these lines lie in the far-UV, observations of H$_2$ in our Galaxy and up to intermediate redshifts require space-based instruments. Pioneering studies with {Copernicus} \citep[e.g.,][]{Savage1977}, and later with {FUSE} \citep[e.g.,][]{Shull2000,Gillmon2006a}, provided a detailed view of molecular gas in the Milky Way and nearby galaxies, while HST/COS extended these studies to $z\sim0.7$ \citep[e.g.,][]{Muzahid2015}.

At higher redshifts, the Lyman-Werner bands of H$_2$ shift into the optical window accessible from the ground  
\citep{Levshakov1985}. In particular, UVES and X-shooter on the Very Large Telescope (VLT) have revealed the presence of H$_2$ in a fraction of damped Ly$\alpha$ systems (DLAs) in quasar spectra at $z\sim2$–3 \citep[e.g.,][]{Ledoux2003, Noterdaeme2008b, Balashev2019}.  
Building on the idea proposed by \citet{Stern2021} that the inner circumgalactic medium in the early Universe may contain a larger neutral fraction than is typically inferred at low redshift, recent simulations with detailed \HI–H$_2$ transition models suggest that such gas can produce DLA-level \HI\ columns while maintaining low molecular fractions  \citep{Gurman2025}. This could explain the low detection rate of H$_2$ in most systems at $z\sim2-3$ although some sightlines with higher atomic and molecular content 
likely intersect galactic discs \citep{Ranjan2018}.

Pushing the observations to $z\sim4$ is key to probe the neutral phase but introduces additional challenges: the increasing density of the Ly$\alpha$ forest, the dimming of background sources, and their declining number all hamper absorption-line studies. 
Fortunately, the recent discovery of bright quasars at \mbox{$z>3$} \citep[e.g.,][]{Boutsia2020, Wolf2020} has enabled a high-resolution ($R \sim 120\,000$) legacy survey of 23 $z$\,$\sim$\,4 quasars with ESPRESSO on the VLT: the QUasar Absorption Line Survey \citep[EQUALS,][]{Berg2025}. 

Here, we report the detection of H$_2$ absorption in a strong DLA at $\zabs \approx 4.24$ toward \JBlong\ (hereafter \JB; $z_Q \approx 4.27$\footnote{{$z_Q=4.26$ was originally derived from the shallow \SiIV\ and \CIV\ emission in the Ir\'en\'ee du Pont telescope discovery spectrum. The presence of \NV\ absorption at $z\approx4.27$ suggest a redshift at or slightly above this value \citep[see][]{Cuellar2025}.}}).
This marks the highest redshift at which H$_2$ absorption has been detected to date. Although only slightly beyond the $\zabs = 4.22$ absorber reported by \citet{Ledoux2006}, its much lower column density would likely have remained undetected without the sensitivity and resolution of our observations. 
{
These are presented together with the data reduction in Sect.~\ref{s:obs}.
The analysis is described in Sect.~\ref{s:analysis}, and the results are presented in Sect.~\ref{s:results}.
We discuss the results and conclude in Sect.~\ref{s:conclusion}.}

\section{Observations and data reduction \label{s:obs}}

\JB\ was observed seven times in October 2023, with each observation consisting of a 3053\,s on-target ESPRESSO exposure. All exposures were obtained in dark time, clear or better conditions with seeing $<0.9\arcsec$, airmass $<1.5$, adopting a $4\times2$ binning in the SINGLEHR mode.
We reduced the data using ESO ESPRESSO pipeline v3.3.10. Individual exposures were shifted to the barycentric frame and combined into a single 1D spectrum using \texttt{Astrocook} \citep{Cupani2020}.
Atmospheric H$_2$O and O$_2$ absorption was modelled in each individual observed exposure using \texttt{Molecfit} \citep{Smette2015} and the resulting transmission spectra were then combined in the same way as the science data to produce an effective model of the atmospheric transmission in the final spectrum.

\section{Analysis \label{s:analysis}}

We performed multi-component Voigt profile fitting to derive the column densities ($N$), redshifts ($z$), and Doppler parameters ($b$) of the \HI, metals and H$_2$ absorption features. 
Telluric absorption lines were taken into account by including the effective atmospheric transmission into the overall fitted model.

{\subsection{Neutral hydrogen \label{a:HImet}}}

\begin{figure*}
    \centering
    \includegraphics[width=\linewidth,clip,trim={3.7cm 0.5cm 2.5cm 0.5cm}]{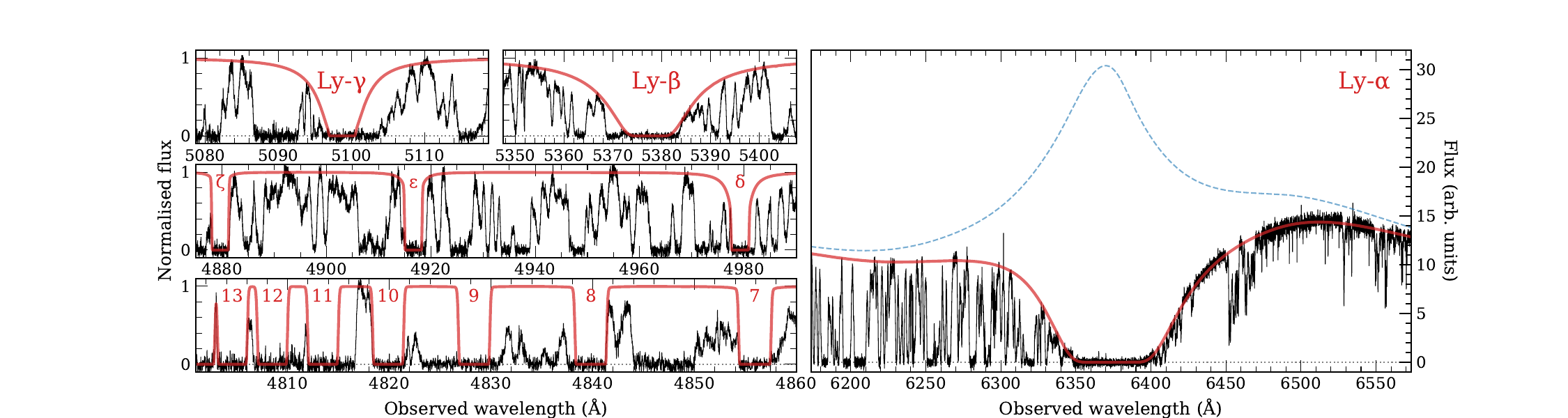}
    \caption{\HI\ Lyman series at $z=4.24$ towards \JB. ESPRESSO data is shown in black, with the synthetic \HI\ profile overplotted in red. 
    For Ly-$\alpha$, the reconstructed quasar emission profile is indicated by the dashed blue line.}
    \label{fig:J0007HI}
\end{figure*}

We initially reconstructed the unabsorbed quasar continuum using a spline function with nodes placed in absorption-free regions. Over the Ly-$\alpha$ emission line 
--affected by the DLA-- we positioned the spline anchor points manually, guided by a matched quasar composite spectrum. The continuum was then refined by simultaneously fitting the Lyman series lines along with a Chebyshev polynomial.
From this procedure, we measure the total neutral hydrogen column density to be 
$\log N(\HI)/{\rm cm}^{-2} = 21.36 \pm 0.05$. The effective Doppler parameter and redshift are also well constrained from the sharp edges of the high order Lyman series to be
be $b_{\HI}^{eff} = 27 \pm 1$\,\kms and $z_{\HI}^{eff} = 4.242708$, see 
Fig.~\ref{fig:J0007HI}.

\subsection{Metals}

\begin{figure*}
    \centering
\includegraphics[width=\linewidth,clip,trim={0.7cm 0cm 0cm 0cm}]{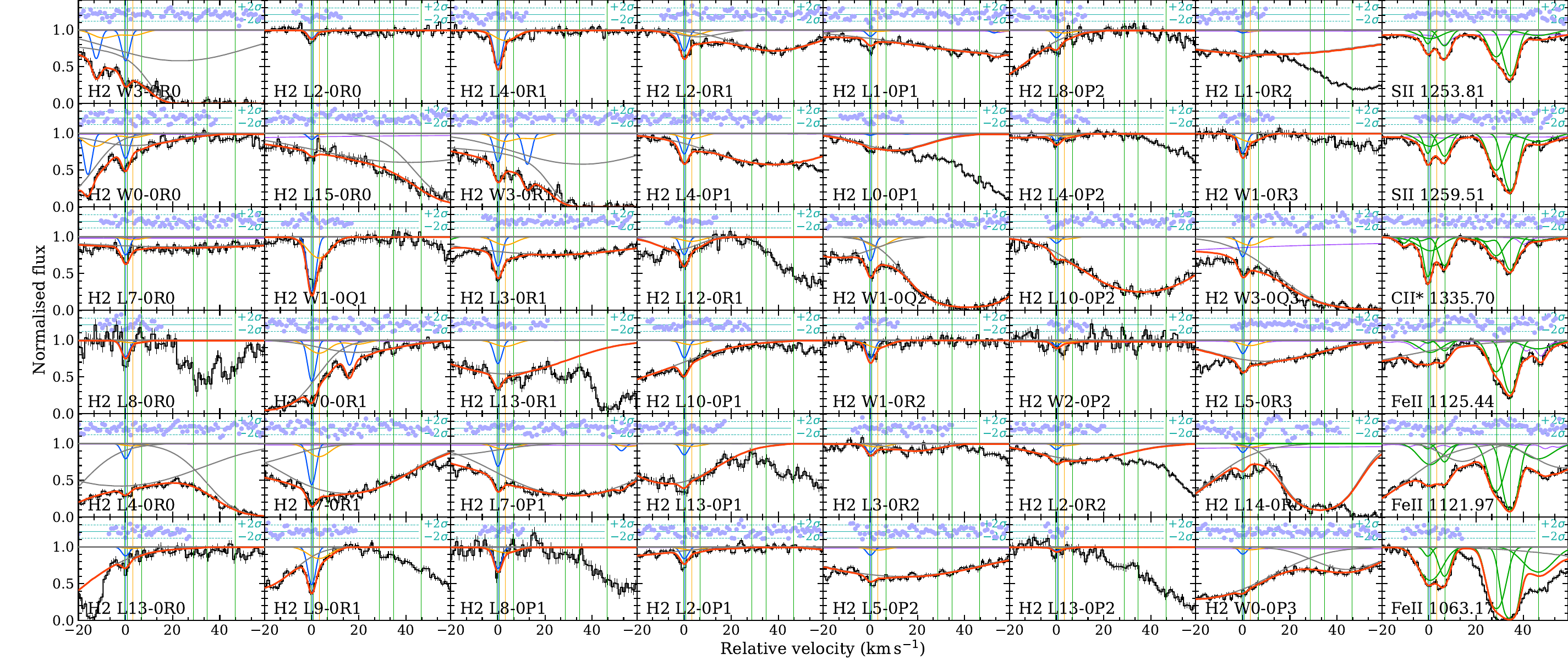}
    \caption{Voigt-profile fit to H$_2$ and metal lines towards \JB. The origin is set at $z_{\HH}=4.242745$. The observed normalised spectrum is shown in black, with the total synthetic absorption in red. H$_2$ components are shown in blue (narrow) and orange (wide) while metal components are shown in green. The transmission from the DLA \HI\ lines, as well as telluric lines (visible in e.g., \SII$\lambda$1253 and \FeII$\lambda$1125 panels), is shown in purple. Additional components in grey, fitted jointly with H$_2$ and metal lines, represent intervening Lyman-$\alpha$ forest lines (the $v\sim20$~\kms component in the H$_2$W3-0R0 panel is actually Ly-$\beta$ from a $z=3.838$ system, also visible in Ly-$\alpha$ at the blue edge of the \FeII$\lambda1121$ panel). }
    \label{fig:J0007}
\end{figure*}

Singly ionised metal absorption lines show that the neutral gas is distributed over only about 50\,\kms\ (Fig.\ref{fig:J0007}), with a shallow satellite component separated by about 60~\kms\ from the rest. Atomic data from \citet{Morton2003} were used for fitting the metals, except for \SII, for which Ritz wavelengths were taken from the NIST Atomic Spectra Database\footnote{\url{https://www.nist.gov/pml/atomic-spectra-database}}. A mismatch with the data
was evident for \SII$\lambda1253$ when using the wavelength from \citet{Morton2003} (1253.805\,\AA) as noted by \citet{Noterdaeme2007b}. In contrast, the Ritz wavelengths provide an excellent match (with 1253.811~{\AA}).
While the profiles of \OI\, \SiII\ and \CII\ are strongly saturated and not fitted, those of \SII, \FeII\ and \CII\ in its first excited level are well reproduced with a seven component model (Table~\ref{t:met}).

\begin{table}
\caption{Results from fitting singly ionised metal lines \label{t:met}}
\addtolength{\tabcolsep}{-0.2em}
\begin{tabular}{ccccc}
\hline \hline
{\large \strut}$z$ & $b$ & \multicolumn{3}{c}{$\log N/\cmsq$}\\ 
& (\kms) & \SII & \FeII & \CII$^*$ \\
\hline 
{\large \strut}4.241769 & $13.0^{+1.0}_{-2.0}$ & $13.59^{+0.04}_{-0.04}$ &  & $11.96^{+0.27}_{-0.28}$ \\
{\large \strut}4.242737 & $1.75^{+0.23}_{-0.22}$ & $13.38^{+0.08}_{-0.08}$ & $12.56^{+0.24}_{-0.24}$ & $12.92^{+0.06}_{-0.05}$ \\
{\large \strut}4.242755 & $6.34^{+0.52}_{-0.52}$ & $13.61^{+0.09}_{-0.10}$ & $13.66^{+0.05}_{-0.05}$ & $12.78^{+0.12}_{-0.13}$ \\
{\large \strut}4.242861 & $2.91^{+0.25}_{-0.25}$ & $13.60^{+0.06}_{-0.07}$ & $13.29^{+0.09}_{-0.10}$ & $12.86^{+0.05}_{-0.06}$ \\
{\large \strut}4.243244 & $4.08^{+0.31}_{-0.30}$ & $13.99^{+0.05}_{-0.06}$ & $13.97^{+0.06}_{-0.05}$ & $12.75^{+0.06}_{-0.06}$ \\
{\large \strut}4.243349 & $3.37^{+0.14}_{-0.15}$ & $14.25^{+0.04}_{-0.03}$ & $14.26^{+0.03}_{-0.04}$ & $12.99^{+0.04}_{-0.04}$ \\
{\large \strut}4.243554 & $8.20^{+0.65}_{-0.65}$ & $13.51^{+0.04}_{-0.05}$ & $13.60^{+0.04}_{-0.03}$ & $12.27^{+0.10}_{-0.10}$ \\
{\large \strut}total & & 14.65$\pm$0.02 &  14.58$\pm$0.03 & 13.60$\pm$0.03 \\
\hline
\end{tabular}
\addtolength{\tabcolsep}{+0.2em}
\end{table}

{\subsection{Molecular hydrogen}}

H$_2$ absorption lines are detected in the first four rotational levels, see  
Fig.~\ref{fig:J0007}.
To account for both continuum uncertainties and frequent blending with intergalactic features, we simultaneously fitted intervening \ion{H}{i} components and H$_2$ lines. Two components --one narrow and one broad-- were required to model the H$_2$ absorption profile. During the fit, the redshifts ($z$) and Doppler parameters ($b$) were tied across rotational levels ($J$). For the narrow component, the Doppler parameter is primarily constrained by the $J=1$ level, which shows numerous transitions spanning from optically thin to intermediate optical depths. In contrast, lines from higher-$J$ levels lie mostly in the optically thin regime, where the column densities are largely insensitive to the $b$-value and the observed line width merely reflects the instrumental line spread function.  The broad component is in turn well resolved 
and its Doppler parameter is directly constrained by the observed line widths.
The optically thin $J\ne1$ lines justify our assumption of a shared $b$ across levels, and, importantly, allows us to distinguish kinematically and thermally distinct components rather than fitting a single, level-dependent effective profile. 
Allowing $b$ to vary with level may actually blend together the contributions of different gas phases, obscuring the underlying structure.

\begin{table}[]
    \centering
    \caption{Results from fitting H$_2$ lines}
    \label{t:H2}
    \begin{tabular}{c c c}
    \hline\hline
        Component:              & narrow    & broad \\
        \hline
        $z$                & 4.242745  &  4.242799 \\
        $b$ (\kms)               & 1.7$\pm$0.1 & 6.2$\pm$0.3 \\
        $\log N($H$_2,J=0)$  & 13.17 $\pm$ 0.04 & 12.75 $\pm$ 0.14\\
        $\log N($H$_2,J=1)$  & 13.81 $\pm$ 0.02 & 13.60 $\pm$ 0.04 \\
        $\log N($H$_2,J=2)$  & 13.23 $\pm$ 0.05 & 13.24 $\pm$ 0.08 \\
        $\log N($H$_2,J=3)$  & 13.28 $\pm$ 0.05 & 13.30 $\pm$ 0.07\\
        total                & 14.06 $\pm$ 0.02 & 13.92 $\pm$ 0.03 \\
        \hline
    \end{tabular}
\end{table}

\section{Results \label{s:results}}

\subsection{Chemical enrichment}

From the total column densities of neutral gas $\log N(\HI)=21.36\pm0.05$ and of volatile singly ionised sulphur, $\log N(\SII)\,=\,14.64\pm0.02$, we infer the average metallicity of the system to be $Z\,\simeq\,0.01\,Z_{\rm \odot}$ and the depletion of iron to be [Fe/S]\,=\,$-0.4$. One component has stronger depletion with [Fe/S]\,$\approx$\,$-1.2$ and coincides (within 0.4~\kms) with the narrow H$_2$ component. This suggests that the main H$_2$-formation route remains catalytic reactions on the surface of dust grains \citep[e.g.,][]{Wakelam2017}, even in this low-metallicity 
environment \citep[see][for a discussion about the possible dominance of gas-phase reactions in environments with low dust content]{Glover2003}.

\subsection{Physical conditions}

The abundance of H$_2$ in the neutral gas depends on the formation-photodissociation equilibrium 
\citep[][]{Jura1975}: 
\begin{equation}
    R\,n\,n_{\HI} = n_{\rm H_2}\,D_0\,\cal{S},  
    \label{form_eq}
\end{equation}
\noindent 
with $n = n_{\HI} + 2n_{\rm H_2}$; $D_0 = 5.8\times10^{-11}I_{\rm UV}\,{\rm s}^{-1}$ where $I_{\rm UV}$ is the UV field intensity in Draine units; $\cal{S}$ is the self-shielding function which, given the low $\rm H_2$ column density, can be approximated here to unity and $R$ is the formation rate onto dust grains.
Introducing the molecular fraction $f_{\rm H_2}=2n_{\rm H_2}/(2n_{\rm H_2}+n_{\HI})$, {Eq.~\ref{form_eq} becomes:}
\begin{equation}
    {n R = D_0 {f_{\rm H_2} \over {2(1-f_{\rm H_2})}}}
    \label{form_eq2}
\end{equation}
Considering the typical density-to-UV ratio in high-$z$ CNM simulations, $\avg{n/I_{\rm UV}} = (N(\HI)/10^{21}\,\cmsq)$\,cm$^{-3}$ \citep{Gurman2025}, the observed $\avg{f_{\rm H_2}}\approx 1.7\times10^{-7}$ implies an average formation rate of $\avg{R} \simeq 2\times10^{-18}$\,s$^{-1}$. %
This is below the standard CNM value in the local ISM \citep[$R \simeq 3\times10^{-17}$\,s$^{-1}$;][]{Jura1975}, but somewhat above expectations at the measured metallicity\footnote{$R<0.4\times10^{-18}$\,s$^{-1}$, accounting for the sub-linear scaling of $R$ with metallicity \citep[e.g.,][]{Roman-Duval2022} and its $\sqrt{T}$-dependence.}.
{In other words, if average densities from the simulations are assumed, reproducing the H$_2$ constraints requires adopting a formation rate higher than implied by the mean metal content. 
This suggests that localised density peaks contribute to H$_2$ production. We note that moderately elevated dust abundances assumptions are also
used in the fiducial model of \citet{Gurman2025} to reproduce global H$_2$ detection rates in DLAs.}

Next, we investigate the regions where H$_2$ is effectively formed. 
From the Doppler parameters of the narrow and broad components ($b=1.7$ and 6.2\,\kms), assuming pure thermal contribution, it is possible to derive strict upper limits on the gas temperatures to be $\sim$350\,K and $\sim$4600\,K. In other words, 
the narrow component arises from cold medium, while the broad could in principle arise from a warmer medium. 

\begin{figure}
       \centering
        \includegraphics[width=\linewidth,clip,trim={0.2cm 0.45cm 0.4cm 0.2cm}]{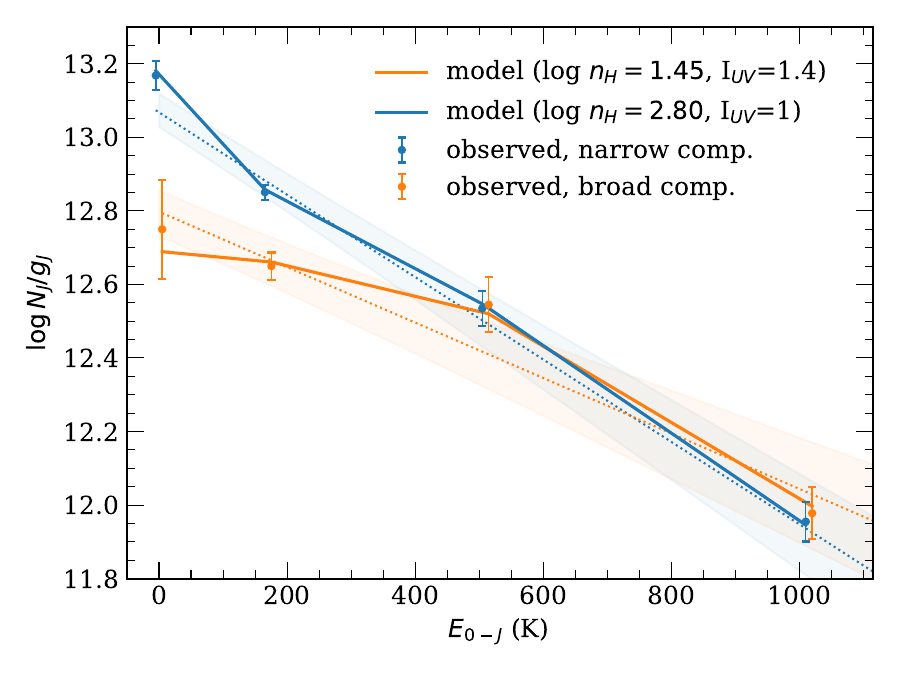}
    \caption{Rotational J=0-3 population of H$_2$ for both  components. Points are artificially shifted by $\pm$5~K along the x-axis for clarity. 
   Dashed lines show single-temperature fits (with shaded uncertainties), and solid lines our best fiducial Cloudy models. 
    \label{f:H2ex}}
\end{figure}

Key information about the gas temperature comes from the population of H$_2$ across different rotational levels (Fig.~\ref{f:H2ex}).
For the narrow component, we measure $T_{01} \approx 233\pm33$\,K, with the distribution across levels ($J=0$–3) reasonably fitted by a single excitation temperature of $T_{\rm ex}\sim390$ K. While $T_{01}$ is generally considered a good proxy for the kinetic temperature of the gas in the high-column-density regime, where H$_2$ is self-shielded in the low rotational levels, it more likely represents an upper limit here, since H$_2$ formation tend to overpopulate $J=1$ relative to $J=0$.
The broad component, in turn, exhibits a flatter excitation diagram with $T_{\rm ex}\sim600\pm100$ K and a $J$=1/$J$=0 ratio consistent with 3 --the ratio of statistical weights-- i.e. the value expected when the ortho–para ratio is set by H$_2$ formation and destruction processes \citep{Abgrall1992}. This directly indicates that the number density in the broad component is lower than in the narrow one, where collisional excitation is likely at play. In contrast, in the broad component, the rotational populations do not directly trace the gas kinetic temperature.
   
To further constrain the physical conditions, we investigated the rotational excitation of H$_2$ in constant-density models with Cloudy v25 \citep[see][]{Chatzikos2023}. We assumed a plane-parallel geometry for the two H$_2$-bearing clouds illuminated by a radiation field. As a starting point, we adopted observational constraints whenever possible and standard values when no constraints were available.
We included a fixed metagalactic background from \citet[][]{Khaire2019} at $z=4.24$ and an additional Draine-like UV field, treated as varying parameter. H-ionising photons were removed, 
as our focus is on the neutral gas. We adopted a Galactic cosmic-ray (CR) ionisation rate of $2\times10^{-16}$~s$^{-1}$ per H$^0$ \citep{Indriolo2007} and scaled the metal and dust abundances to 1\% the local ISM values, according to the mean DLA metallicity.

\begin{figure}
    \centering
        \includegraphics[width=\linewidth]{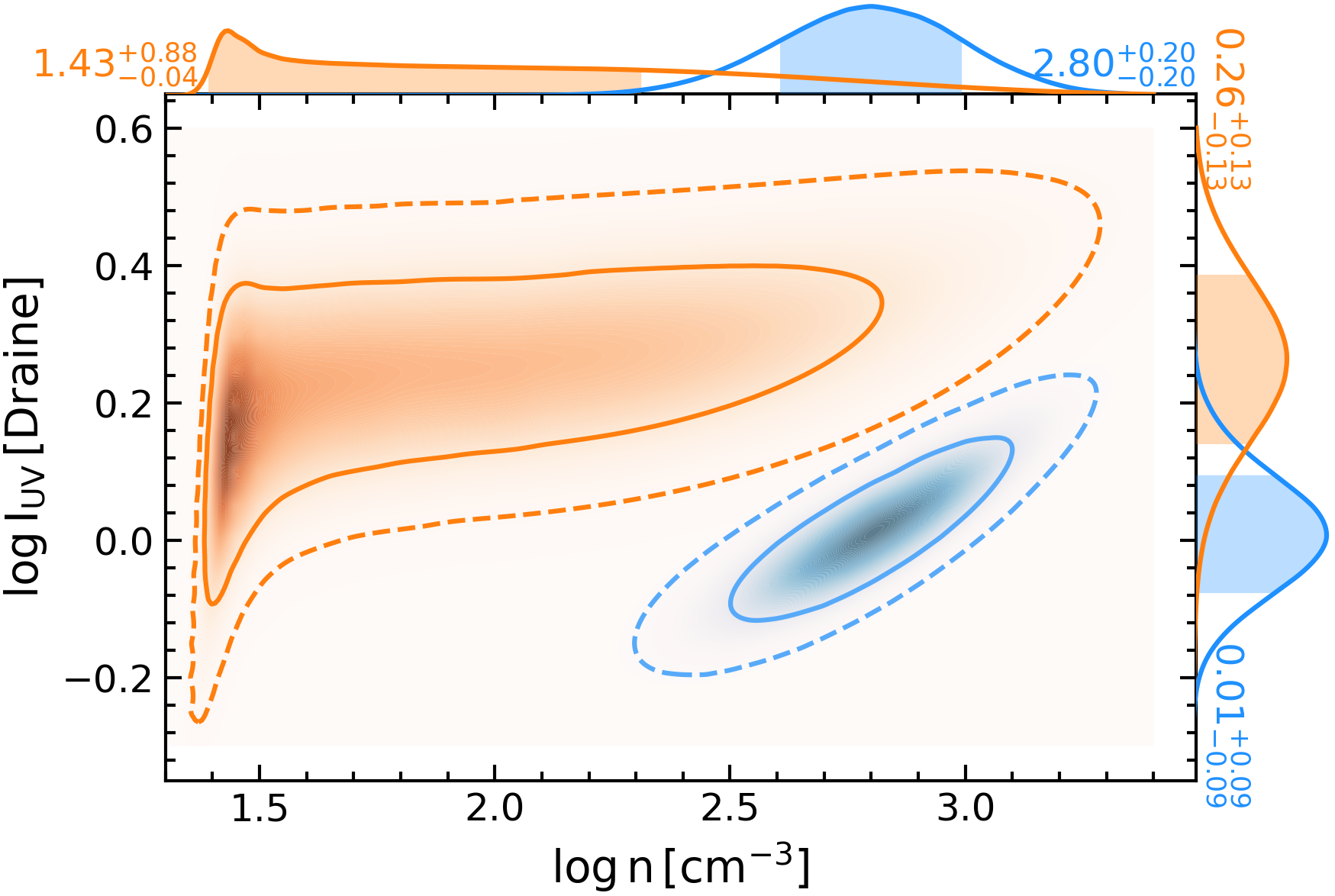}
    \caption{Posterior distributions of the UV field and number density in the broad (orange) and narrow (blue) components. The likelihood was derived by comparing the observed H$_2$ J=0 to J=3 column densities with those from a grid of Cloudy models and converted into 2D posterior distribution assuming independent flat priors in logarithmic space. The 1D marginalised posteriors are shown along the top and right axis.
    \label{f:cloudygrid}
    }
    \end{figure}

Under these assumptions, we created model grids varying only the gas density and the strength of the UV field. Our 
fiducial model reproduces best the observed H$_2$ excitation in both components for densities of $\log n_{\rm H} \approx 2.8$ 
and 1.4 for the narrow and broad components, respectively, under a UV field comparable with the Galactic level for both components, 
 see Figs.~\ref{f:H2ex} and \ref{f:cloudygrid}.
The modelled gas temperatures, 40 and 560~K correspond to similar thermal pressure ($P_{\rm th}/k_{\rm B}\sim 2\times 10^4$\,K\,cm$^{-3}$) in the two components and are consistent with the limits implied by the Doppler parameters.
Although we did not attempt to reproduce the metal column densities (as the association between metals and H$_2$ components is uncertain and the depletion pattern is not well constrained), the predicted \SiII\ and \CII*\ columns are reasonably close to (below) the observed values, providing additional confidence in the model.  

We also tested models with CR ionisation rates varied by a factor of 10 and with either baseline or tenfold higher metal and grain abundances. These provide poorer fits to the H$_2$ excitation and metal column densities, especially when assuming identical UV field and CR rate for both components.
{A more realistic model, where the cold component has a higher gas-phase metallicity—consistent with stronger iron depletion—but still a low grain abundance (as dust scales sub-linearly with metals), improves the match to the \CII* and \SiII\ columns without changing the derived density or UV field.}
Since \CII*\ excitation scales as $nT^{0.35}$ \citep{Balashev2022}, the derived temperatures and densities imply a \CII$^*$/\CII\ ratio about four times higher in the narrow component, in agreement with the observations.

\subsection{Length scales}

The inferred densities imply length scales of $L \simeq 0.01$\,pc (narrow H$_2$ component) to a few pc (broad component), pointing to a highly fragmented neutral medium, consistent with lower-redshift studies of self-shielded H$_2$ systems \citep[e.g.,][]{Jorgenson2009,Balashev2011}.
Here, H$_2$ is merely a tracer of cold gas yet underlining it is structured to very small scales. 
That the larger cloud is also the more turbulent one is  
consistent with the trend expected from Larson’s scaling 
\citep[$b_{turb} \propto L^{0.38}$, ][]{Larson1981}, 
providing a self-consistent link between cloud sizes and the underlying turbulent dynamics of the gas.
The coexistence of these narrow and broad H$_2$ components likely reflects the underlying 
$n_{\rm H}$ distribution shaped by turbulence and self-gravity. Such a close physical association could also explain the previously reported increase in $b$-parameter when the absorption is considered as a whole \citep[e.g.,][]{Jenkins1997,Noterdaeme2007b,Tchernysyhov2022}. 

{Motivated by the very small inferred size of the cold cloud, we investigate the possibility of partial coverage.} This occurs when the absorbing cloud does not fully cover the background source, which is expected when the cloud’s characteristic size is smaller than, or comparable to, that of the emitter at the absorption wavelength. This effect is most readily identified through a non-zero residual flux at the bottom of saturated lines, as observed, for example, in strong H$_2$ absorption against the parsec-scale broad emission-line regions of quasars \citep[e.g.,][]{Balashev2011}.

\begin{figure}
    \centering
    \includegraphics[width=1.0\linewidth]{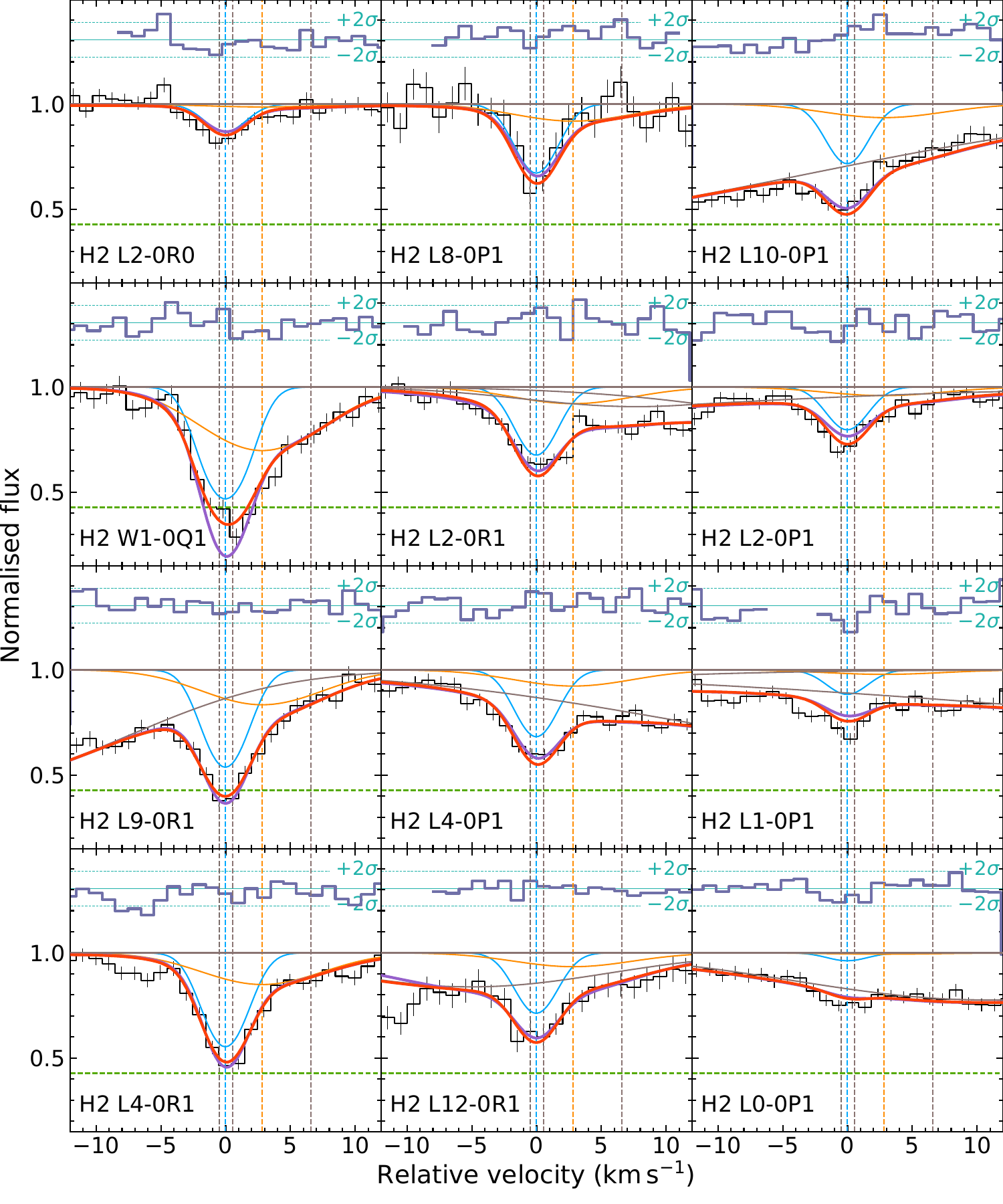}
\caption{Comparison of the best-fit model including partial coverage for the narrow component (red) and assuming full coverage (violet). Only a subset of H$_2$ lines is shown, focusing on transitions where profile differences are most pronounced. Graphical elements follow Fig.~\ref{fig:J0007}. {The green dashed horizontal line depicts $(1-C_f)$, the fraction of the background source uncovered by the narrow H$_2$ component}.}
    \label{f:Cf}
\end{figure}

In the present case, the 0.01-pc length scale inferred for the cold component is comparable to the light-day scale of quasar accretion discs \citep[e.g.,][]{Yu2020}, against which the H$_2$ absorption is observed. However, the H$_2$ lines are optically thin so any partial-coverage signature is expected to be subtle. Nevertheless, introducing the covering fraction as a free parameter {for the narrow H$_2$ component} yields a covering factor of $C_f \sim 60\%$ and a slight improvement in the fit, most noticeably for a few H$_2$ lines (Fig.~\ref{f:Cf}) and 
providing an independent, transverse indication of the small cloud size. 
{While the formal preference of the model with partial covering factor is high ($\Delta \rm AIC\approx 45$, where AIC is the Akaike information criterion), the improvement is modest in a visual sense\footnote{We checked that it is not only driven by the strongest transition, W1-0Q1, see Fig.~\ref{f:Cf}}. Higher signal-to-noise data will be required to robustly confirm partial coverage and to fully disentangle it from possible modelling systematics that could arise from component decomposition, line spread function, local continuum placement, or local deviation in noise properties.}

In this scenario, the inferred column densities in all rotational levels increase by $\approx$0.3~dex (Table~\ref{t:Cf}), with no noticeable impact on the derived physical conditions. The inferred cloud size becomes about twice as large, and the predicted metal column densities increase accordingly, bringing them closer to the observed values.

\begin{table}
    \centering
    \caption{Results from fitting H$_2$ lines allowing for partial coverage for the narrow component}
    \label{t:Cf}
    \begin{tabular}{c c c}
    \hline\hline
        component              & narrow    & broad \\
        \hline
{\large \strut}                $z$                & $4.2427455(1)$  &   $4.242795(7)$ \\

{\large \strut}        $b$ (\kms)               & $1.59\pm0.12$ & $6.2\pm0.4$ \\
{\large \strut}$\log N($H$_2,J=0)$ & $13.49\pm0.06$ & $12.67\pm0.23$  \\
{\large \strut}$\log N($H$_2,J=1)$ & $14.20\pm0.06$ & $13.63\pm0.04$  \\
{\large \strut}$\log N($H$_2,J=2)$ & $13.52\pm0.06$ & $13.25\pm0.09$   \\
{\large \strut}$\log N($H$_2,J=3)$ & $13.59\pm0.07$ & $13.31\pm0.09$  \\
\hline
{\large \strut}$C_f$ & $0.57\pm0.04$ & \\
        \hline
    \end{tabular}
\end{table}

\section{Discussions and conclusion \label{s:conclusion}}

As the inferred incident UV radiation field is close to the Galactic value ($I_{\rm UV} \simeq 1$), the absorber likely lies far from the quasar influence. Given the quasar’s brightness ($i = 17.2$) and redshift ($z \simeq 4.27$), its UV luminosity ($L_\nu \sim 3\times10^{32}$~erg~s$^{-1}$~Hz$^{-1}$) would produce a Galactic-level field only beyond $\sim$2~Mpc, consistent with the $\sim$1500~\kms\ velocity offset between the quasar and the DLA ($\sim$3~Mpc for pure Hubble flow).  This illustrates that proximity in velocity space does not necessarily imply physical proximity. Conversely, the inferred UV field can be explained by the quasar alone,  without requiring additional sources. 
This, in turn, suggests that the absorbing gas may reside {far away} from any luminous galaxy. 

{The physical conditions inferred for the H$_2$-bearing gas are broadly consistent with those reported for other H$_2$-bearing DLAs \citep[e.g.,][]{Petitjean2000,Cui2005,Noterdaeme2007a,Jorgenson2010,Albornoz2014,Noterdaeme2017,Balashev2019,Klimenko2020,Balashev2024}, as well as for diffuse molecular gas observed along high-latitude lines of sight through the Milky Way disc and halo \citep{Richter2003,Gillmon2006a,Wakker2006}. However, the gas densities inferred here—particularly for the narrow component—lie toward the upper end of the range typically reported in previous studies. This relatively high density allows trace amounts of H$_2$ to form in a low-metallicity environment without self-shielding. In this optically thin regime, the gas density determines both the kinetic temperature and the H$_2$ excitation. By contrast, the broader component has a density closer to those commonly inferred in the literature, although such densities are usually associated with much higher molecular columns; its temperature is also substantially higher. What distinguishes the present system, therefore, is not extreme physical conditions per se, but the presence of dense, although almost fully atomic gas.}

Even in the cold clump, the molecular fraction remains {indeed} diminute ($\log f_{\rm H_2} < -5$), highlighting H$_2$ as a sensitive overdensity tracer 
in low-metallicity neutral gas. 
Here, it could trace a very compact structure, with a density of $\log n_{\rm H} \sim 2.8$ {over a length scale} of only $\sim 0.01$~pc. 
Yet, the detection of such {an absorber} among only seven DLAs in the EQUALS sample suggests that these {overdensities} may be relatively common.
This system also likely reflects key evolutionary effects: a higher average gas density at these redshifts and a lower metallicity. The latter reshapes the thermal balance, restricting the formation of cold gas to higher densities because of the reduced metal-line cooling.

A second example is found at $z=2.97$ in the EQUALS spectrum of J\,0204–3251 (Appendix \ref{s:J0204}). Despite its lower S/N and more uncertain measurements, this system presents similar properties: 
a relatively low H$_2$ column density distributed over two closely spaced components and a broader component ($b\sim2.6$~\kms) more highly excited than the narrow one ($b\sim0.8$~\kms).
Cold overdensities could then actually be frequent enough in the neutral medium at high redshift 
to collectively build a significant covering fraction.

Individual CNM cloudlets would be hardly detectable in 21-cm absorption, even with the Square Kilometer Array, because their physical sizes are much smaller than the extent of 
radio sources. A sufficiently large covering factor could in principle make their collective 21-cm imprint detectable, despite their low volume filling factor, but the signal could remain shallow, as clouds may spread over a relatively broad velocity interval.
In the optical, such compact structures are also challenging to identify at lower spectral resolution or signal-to-noise than achieved here. Their detection in this system was possible only thanks to the achieved resolving power 
combined with the brightness of the background quasar. Even the broader $\sim 4$-pc component would have remained hidden without the narrow feature flagging the presence of cold gas. ESPRESSO’s sensitivity to narrow absorption lines thus provides a rare probe of the small-scale density structure of the neutral medium at high redshift.
Looking ahead, using weak H$_2$ as a routine tracer of CNM cloudlets in low metallicity gas at high redshift \citep[potentially leading to star formation before reaching high molecular fraction;][]{Krumholz2012,Glover2012}, will require much larger collecting areas, as will be provided by ANDES on the ELT. 
Re-observing known H$_2$-bearing systems at higher spectral resolution will also help isolate multiphase structures and uncover weak H$_2$ components that are otherwise difficult to detect.

\begin{acknowledgements}
{We thank the referee for useful comments and suggestions.}
We thank IFPU in Trieste for hospitality. SL acknowledges support by FONDECYT grant 1231187.
\end{acknowledgements}

\vspace{-0.5cm}
\bibliographystyle{aa} 
\bibliography{mybib}

\begin{appendix}

\section{H$_2$ at $z\sim3$ towards J\,0204-3251 \label{s:J0204}}

\begin{figure}[!h]
    \centering
    \includegraphics[width=\hsize]{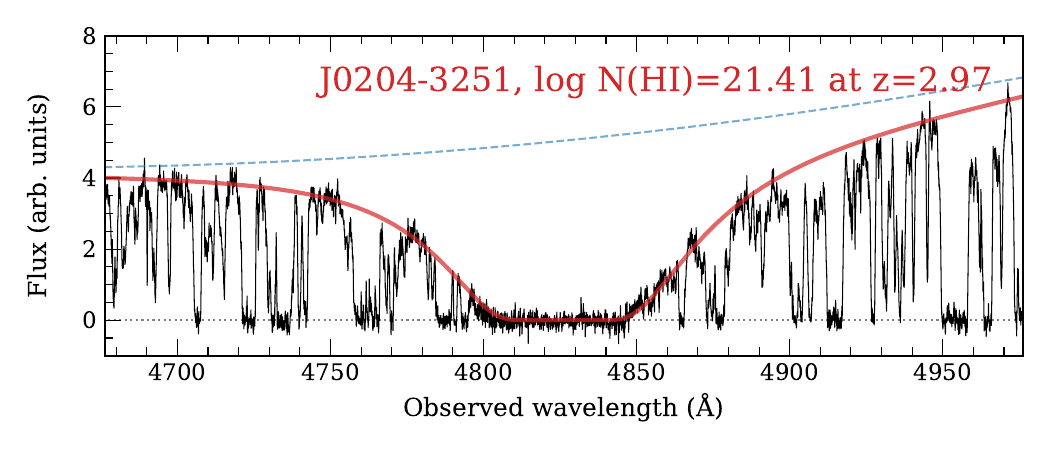}
    \includegraphics[width=\hsize,trim={0 0.6cm 0 0},clip=]{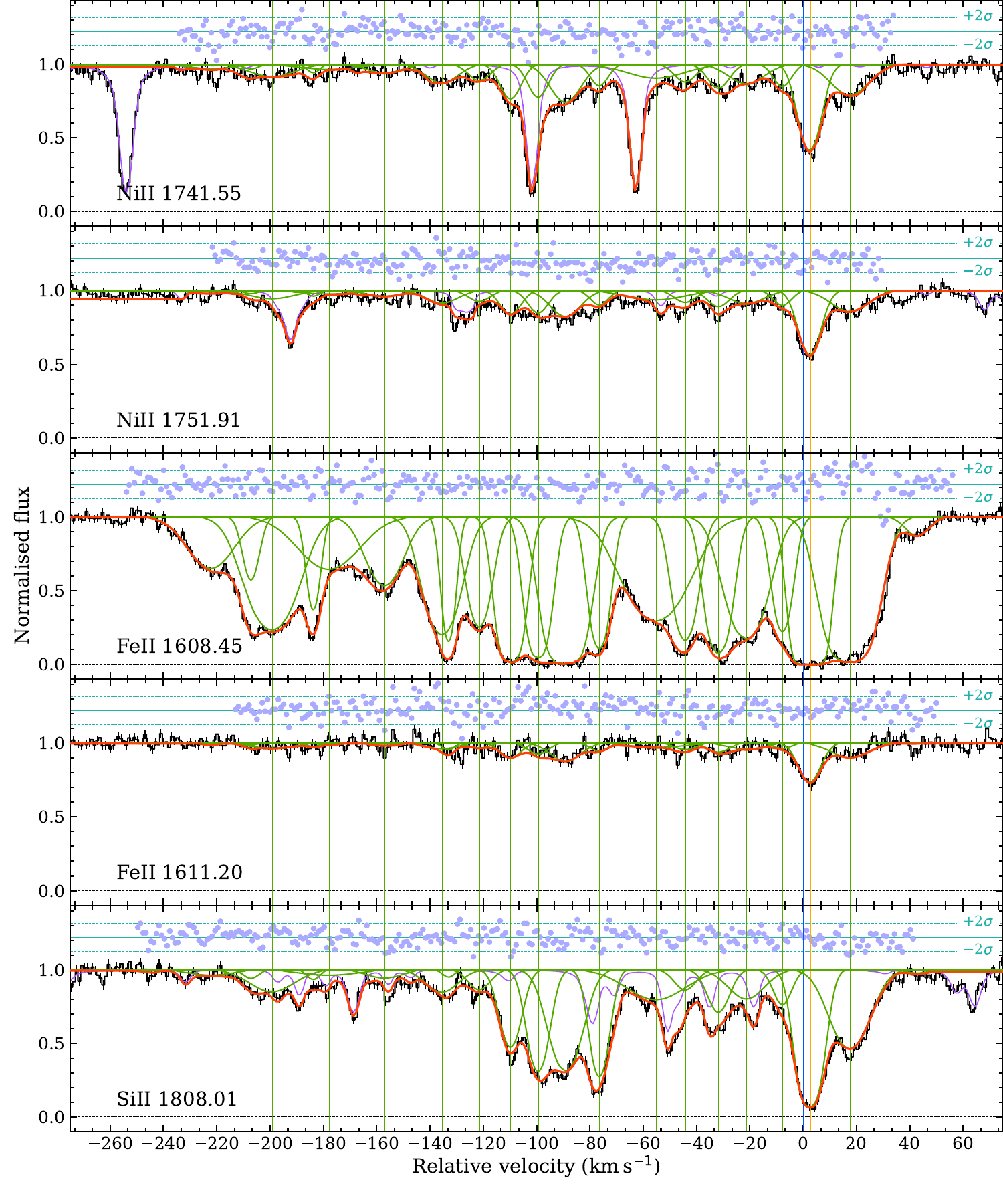}
    \caption{Fit to \HI\ and low-ionisation metals in the $z=2.97$ DLA system towards J0204-3251. The zero of the velocity scale corresponds to the strongest H$_2$ component. Colours are as per Fig.~\ref{fig:J0007}}
    \label{J0204:HI}
\end{figure}

The quasar J\,0204-3251 ($z_Q\simeq 3.8$) has also been observed with ESPRESSO as part of the EQUALS survey. It features a strong intervening DLA at $\zabs \simeq 2.97$ with $\log N(\mathrm{H,I}) = 21.40 \pm 0.01$. The metal-line profile is complex, with numerous components spanning $\sim 250$\,\kms, and indicates an overall metallicity of [Si/H]~$\approx -0.86$ (Fig.~\ref{J0204:HI}). Because silicon is mildly depleted, this value likely represents a lower limit to the true metallicity; \ZnII\ lines are not covered and \SII\ lines fall within the Ly$\alpha$ forest.

\begin{figure}[!h]
    \centering
    \includegraphics[width=\hsize,trim={0cm 0cm 0 0},clip=]{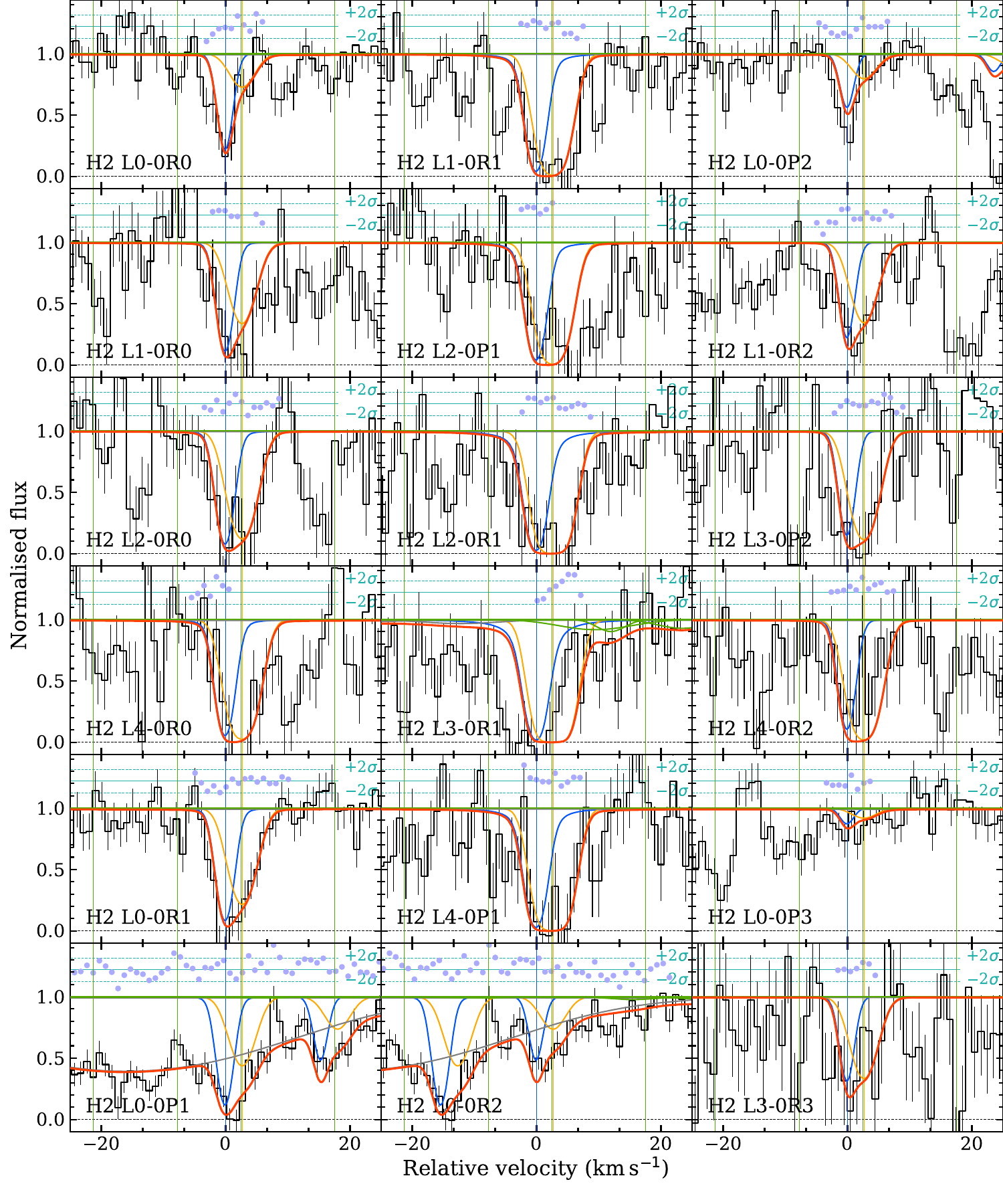}
    \caption{H$_2$ absorption lines towards J\,0204$-$3251}
    \label{J0204:H2}
\end{figure}

\begin{figure}[!h]
    \centering
    \includegraphics[width=0.95\hsize,trim={0.2cm 0.4cm 0.4cm 0.2cm},clip=]{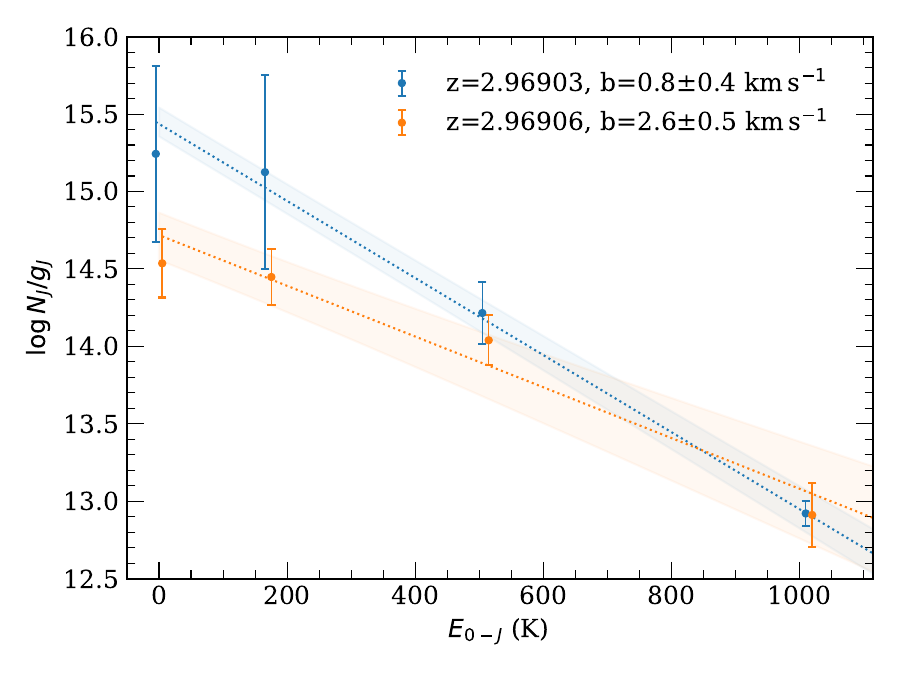}
    \caption{H$_2$ excitation diagram at $z=2.97$ towards J\,0204$-$3251}
    \label{J0204ex}
\end{figure}

Although this spectrum has a lower S/N (particularly in the blue) and covers fewer H$_2$ bands than the main system discussed in this paper, the higher H$_2$ column density enables the detection of two closely spaced components ($\Delta v \approx 2.8$\,\kms) from $J=0$ to $J=3$ (Fig.~\ref{J0204:H2} and Table~\ref{t:J0204}). The corresponding excitation diagram is shown in Fig.~\ref{J0204ex}.

\begin{table}[!h]
    \centering
    \caption{Results from fitting H$_2$ lines towards J\,0204$-$3251.}
    \label{t:J0204}
    \begin{tabular}{c c c}
    \hline\hline
        $z$                & 2.969028 &  2.969063\\
        \hline
        $b$ (\kms)               & 0.8$\pm$0.4 & 2.6$\pm$0.5 \\
        $\log N($H$_2,J=0)$  & 15.2 $\pm$ 0.6 & 14.5 $\pm$ 0.2\\
        $\log N($H$_2,J=1)$  & 16.1 $\pm$ 0.6 & 15.4 $\pm$ 0.2 \\
        $\log N($H$_2,J=2)$  & 14.9 $\pm$ 0.2 & 14.7 $\pm$ 0.2 \\
        $\log N($H$_2,J=3)$  & 14.2 $\pm$ 0.1 & 14.2 $\pm$ 0.2\\
        total                & 16.2 $\pm$ 0.5 & 15.6 $\pm$ 0.1 \\
        \hline
    \end{tabular}
\end{table}

\end{appendix}
\end{document}